\documentclass[12pt]{article}
\usepackage{graphicx,amsmath,amssymb,url,enumerate,mathrsfs,epsfig,color}
\usepackage{tikz}
\usepackage{float}

\usepackage{amsmath}
\usepackage{amssymb}
\usepackage{yfonts}

\usepackage{tikz,pgfplots}

\usetikzlibrary{calc, patterns,arrows, shapes.geometric}

\usepackage{graphicx,amsmath,amssymb,url,enumerate,mathrsfs,epsfig,color}

\usetikzlibrary{decorations.text}
\usetikzlibrary{decorations.markings}

\pgfplotsset{compat=1.8}

\usepackage{enumerate}
\usepackage{boxedminipage}
\usepackage{makeidx}
\usepackage{multicol}
\usepackage{xcolor}
\usepackage{mathtools}

\usetikzlibrary{calc, patterns,arrows, shapes.geometric}

\newenvironment{proof}{\noindent{\bf Proof~:}}{\QED\medskip}
\def\QED{\hskip0.1em\hfill\null\ \null\nobreak\hfill
\kern3pt\lower1.8pt\vbox{\hrule\hbox
{\vrule\kern1pt\vbox{\kern1.7pt \hbox{$\scriptstyle
QED$}\kern0.2pt}\kern1pt\vrule}\hrule}}

\newcommand{\hor}{\mathrlap{\perp}\square}
\newcommand{\ram}{\raisebox{0.22ex}{$-$}}
\newcommand{\ver}{{\mathrlap{\ram}}\square}

\newtheorem{prop}{Proposition}[section]

\newtheorem{definition}{Definition}[section]

\newtheorem{example}{Example}[section]
\title{Material geometry of binary composites}
\author{ Marcelo Epstein \\ \small{Department of Mechanical and Manufacturing Engineering} \\ \small{ University of Calgary, Canada. Email: mepstein@ucalgary.ca}}

\date{}
\begin{document}
\maketitle

{\bf Abstract}: The constitutive characterization of the uniformity and homogeneity of binary elastic composites is presented in terms of a combination of the material groupoids of the individual constituents. The incorporation of these two groupoids within a single double groupoid is proposed as a viable geometric differential framework for a unified formulation of this and similar kinds of problems in continuum mechanics.

\bigskip


\section{Introduction}

Material geometry is the application of differential-geometric methods to the representation of defects in continuous media. Its origins can be traced back to a long memoir published in 1907 by Vito Volterra \cite{volterra}, in which the notion of dislocations (which Volterra originally called distorsions) was first defined. These dislocations, obtained by a process of cutting and welding in different ways a hollow cylinder, give rise to the appearance of strains  without the need for sustained external forces. Later, researchers began to refer to these as {\it eigenstrains},\footnote{This hybrid German-English term has its origins in a pioneering article by Hans Reissner \cite{reissner}, where the German neologism {\it Eigenspannung}, also {\it Selbstspannung}, meaning proper strain or self-strain, was introduced.} whose presence can be regarded as endowing the body with a Riemannian metric of non-vanishing curvature. A rigorous modern treatment that obtains exact solutions for the stress fields produced by eigenstrains in the context of nonlinear elasticity was developed by Yavari and Goriely \cite{arash}.

The modeling of dislocations in continua, whether distributed or localized, is not necessarily tied to eigenstrains or similar concepts. Indeed, it is possible to undertake what may be called purely {\it structural} formulations in which the defects are treated as independent geometric entities. These formulations can be heuristically related to atomic considerations appealing to physicists, but still expressed in differential geometric and topological languages. A purely geometric theory that unifies the treatment of smooth and singular defects has been formulated \cite{reuven1} in terms of differential forms and their distributional generalization known as de-Rham currents. Smectic liquid crystals, in which the molecules follow planar arrangements, and nanotube composites are particularly amenable to this treatment \cite{dual}.

At the other end of the spectrum, one finds the purely {\it constitutive} approach, with its uncompromising adherence to the tenets of continuum mechanics. Indeed, in this discipline it is epistemologically agreed that {\it all} the information pertaining to the material response is encapsulated in the constitutive equations. Accordingly, the existence of `dislocations' is interpreted as violations of the integrability of certain geometric structures that must necessarily emerge from the constitutive laws themselves and from nowhere else. This approach, spearheaded by Walter Noll \cite{noll}, has as its essential building block the so-called {\it material diffeomorphism} between pairs of points in the body, namely, a map between the respective tangent spaces that renders the constitutive responses identical. This {\it transplant} procedure abides by an automatic multiplicative decompostion, an operation that emerges naturally from the rule of transformation of constitutive laws under a change of reference configuration. From this point of view, the so-called `intermediate configuration' is a fictitious and completely unnecessary entity.

In Noll's terminology, a body is said to be {\it materially uniform} if all its points are mutually materially isomorphic. In a uniform body, a smooth field of material isomorphisms with a fixed archetype is nothing but a distant parallelism. If the torsion of the associated connection vanishes, the body can be brought to a homogeneous configuration and is, thus, free of obstacles (dislocations). This analysis applies when the symmetry group of the material is discrete. When it is not, such as in the case of isotropic or transversely isotropic solids, the material connection is not unique and the obstacle to integrability is different for each continuous symmetry group. These and other important issues were tackled by C-C Wang \cite{wang}, who expressed his results in terms of reductions of the principal bundle of frames of the body manifold. In this way, the theory of G-structures made its appearance \cite{sniat} as the next differential geometric object of interest within the purely constitutive approach.

The passage from G-strctures to groupoids \cite{jgp} was expected and welcomed, and for good reasons. When a point in the base manifold is arbitrarily selected, a transitive Lie groupoid can be boiled down to a principal bundle. A different choice leads to an equivalent principal bundle, in a manner reminiscent to the relation between an affine space and the vector space that is obtained after arbitrarily choosing an origin. But more important than this detail is the fact that G-structures can only model uniform bodies, whereas to every material body, whether uniform or not, one can uniquely assign a material groupoid. Transitivity of the material groupoid corresponds to uniformity, while the lack of transitivity can be further exploited to subdivide the body into transitivity components of possibly different dimensions. Thus, a laminated body is not uniform, but its material groupoid still displays a clear geometric structure that leads to physically meaningful characterizations of inhomogeneity. The Lie algebroid associated with a Lie groupoid is a kind of infinitesimal version of the latter and carries information pertinent to the determination of the presence of inhomogeneities \cite{algebroid1}. Somewhat surprisingly, even when the material groupoid is only an algebraic (rather than a Lie) subgroupoid of the 1-jet groupoid of the body, a {\it singular material distribution} can be defined \cite{victor, victor1} and used to obtain a material foliation with transitive leaves representing smoothly uniform sub-bodies, layers, filaments, and isolated points.

The progression towards sharper geometrical tools that are increasingly adapted to the higher complexity exhibited by materials and their uniformity and homogeneity characterization is still ongoing. The aim of this paper is to suggest that the modeling of composite materials, and possibly other continuum mechanics entitites, may be better accommodated within the category of {\it double groupoids}. These are particular cases of {\it double categories} and were introduced into the mathematical literature by Charles Ehresmann \cite{ehresmann}. To keep the presentation as self-contained as possible, Section \ref{sec:groupoids} is a fairly detailed, though elementary, introduction to groupoids and their conceptual relation to the notions of local and distant symmetries in physics. In Section \ref{sec:material}, after a brief introduction to the basic kinematic and constitutive elements of elasticity theory, the definition of the material groupoid is briefly reviewed. Section \ref{sec:doublegroupoid} is entirely devoted to double groupoids. Following an introduction to the topic, an application of the theory to the description of binary composites is proposed.

\section{Groupoids}
\label{sec:groupoids}

\subsection{Definition}

The {\it groupoid} is somewhat of a latecomer to the mathematical repertoire, its earliest appearance and naming being traceable to an article \cite{brandt} by Heinrich Brandt (1886-1954) published in 1927 in the prestigious Mathematische Annalen.\footnote{The adjective `prestigious' is not being applied lightly. A glance at the fist page of Volume 96, in which Brandt's article appears, reveals that the journal had been founded in 1868 by Alfred Clebsch (1833-1872) and Carl Neumann (1832-1925), both important contributors to mathematics and applied mechanics. Carl Neumann is best remembered for his formulation of a type of boundary conditions in elliptic partial differential equations. But the list of editors is even more impressive, including Felix Klein, David Hilbert, Albert Einstein, Constantin Carath\'eodory, Richard Courant, Theodor von K\'arm\'an, and Arnold Sommerfeld.} It is quite remarkable that its author provides, at the start of his short paper, a definition which is as clear as a mathematical definition can be, without indulging in some of the more recent stylistic practices. It is clear that Brandt had an original idea, and that he was eager to communicate it. The very title of Brandt's paper hints at his basic idea and choice of terminology: {\it ``On a generalization of the concept of group''}. Brandt's definition consists of introducing in a set, just as in the case of a group, a binary internal operation (the {\it product} or {\it composition}), but (and this is the essential difference between a group and a groupoid) this operation is {\it not necessarily defined for all ordered pairs} of elements. In other words, given two elements $a, b$, the composition $c=ab$ may, or may not, exist. This operation, however, is not arbitrary, but must satisfy a number of algebraic properties (such as associativity and existence of inverses and some special elements which act as right and left units of sorts). 

 Brandt's definition, formulated in the spirit of an algebraic generalization of the concept of group, can be replaced with an equivalent one which cleverly describes the elements of a groupoid as {\it arrows} between points of an underlying set. This definition, with its clear geometric and physical overtones, turns out to be a creative tool for the discovery of applications, such as the ones intended in this paper. Let us, therefore, consider a {\it total set} $\mathcal Z$, whose elements are the arrows, as our point of departure. Each arrow $z \in {\mathcal Z}$ has a tail (or {\it source}) and a head (or {\it target}), both of which will be considered as elements of another set, the set of {\it objects} $\mathcal B$.
 
\begin{figure}[H]
\begin{center}
\begin{tikzpicture} [scale=1.0]
\tikzset{->-/.style={decoration={
  markings,
  mark=at position .75 with {\arrow{stealth'}}},postaction={decorate}}}
  \shade[ball color=blue!20] (2.4,5.25) ellipse (3.5 and 1.75);
\node at (-0.5,5.25) {$\mathcal Z$};

\begin{scope} [yshift=-30]
\draw[fill=olive!40] (-1.,2) to [bend left=10] (5,2) to [bend left] (6,3) to [bend right=10] (0.,3.25) to [bend right] (-1.,2);
\node at (-0.5,2.4) {$\mathcal B$};
\draw[thick, -stealth'] (0.5,6.6) to [bend left] (3.5,6);
\node at (2,7) {$z$};
\draw[dashed,->-, -o] (0.5,6.6) -- (0.5,2.6);
\node[left] at (2.5,2.6) {$X=\alpha(z)$};
\draw[dashed,->-, -o] (3.5,6.) -- (3.5,3);
\node[right] at (3.5,2.9) {$Y=\beta(z)$};
\end{scope}
\end{tikzpicture}
\end{center}
\caption{A groupoid $\mathcal Z \rightrightarrows \mathcal B$ as a cloud of arrows hovering over a meadow $\mathcal B$}
\label{fig:cloud1}
\end{figure}
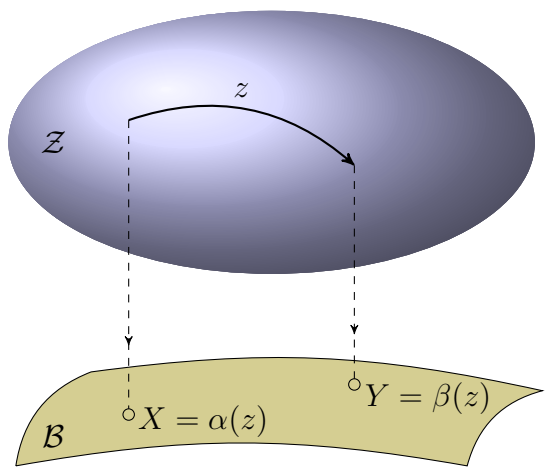

 Pictorially, we may imagine the total set $\mathcal Z$ as a cloud of arrows hovering in place above a meadow $\mathcal B$. These two sets, $\mathcal Z$ and $\mathcal B$, are linked by means of two {\it projection} maps, $\alpha: {\mathcal Z} \to {\mathcal B}$ and $\beta:{\mathcal Z} \to {\mathcal B}$, called, respectively, the {\it source map} and the {\it target map}. As indicated by their names, $\alpha(z)$ and $\beta(z)$ give us, respectively, the tail $X$ and the tip $Y$ of the arrow $z$. This is shown schematically in Figure \ref{fig:cloud1}, but it is important to take this representation with a grain of salt. Arrows, tips, and tails, are just figures of speech. Thus, an arrow does not quite have intermediate points, and the tip and the tail of an arrow in $\mathcal Z$ belong to the set $\mathcal B$. All that the arrow imagery conveys is that there is a certain `relation' between $X$ and $Y$. An elegant and suggestive notation for a groupoid is ${\mathcal Z} \rightrightarrows {\mathcal B}$.
 
\begin{figure}[H]
\begin{center}
\begin{tikzpicture} [scale=1.0]
\tikzset{->-/.style={decoration={
  markings,
  mark=at position .35 with {\arrow{stealth'}}},postaction={decorate}}}
  \shade[ball color=blue!20] (2.4,5.25) ellipse (3.5 and 1.75);
\node at (-0.5,5.25) {$\mathcal Z$};

\begin{scope} [yshift=-30]
\draw[fill=olive!40] (-1.,2) to [bend left=10] (5,2) to [bend left] (6,3) to [bend right=10] (0.,3.25) to [bend right] (-1.,2);
\node at (-0.5,2.4) {$\mathcal B$};
\draw[thick, -stealth'] (0.5,6.6) to [out=180, in=180] (0.5,7.5) to [out=0,in=0] (0.55,6.6);
\node at (1.85,7.3) {$id_X=\epsilon(X)$};
\draw[dashed,->-, o-]   (0.5,2.6)--(0.5,6.6);
\node[right] at (0.5,2.6) {$X$};
\draw[thick, -stealth'] (3.5,6) to [out=180, in=180] (3.5,6.9) to [out=0,in=0] (3.55,6.);
\node at (4.9,6.4) {$id_Y=\epsilon(Y)$};
\draw[dashed,->-, o-]   (3.5,3)--(3.5,6.);
\node[right] at (3.5,2.9) {$Y$};
\end{scope}
\end{tikzpicture}
\end{center}
\caption{The object inclusion map $\epsilon: {\mathcal B} \to {\mathcal Z}$ and the identities}
\label{fig:identities}
\end{figure}
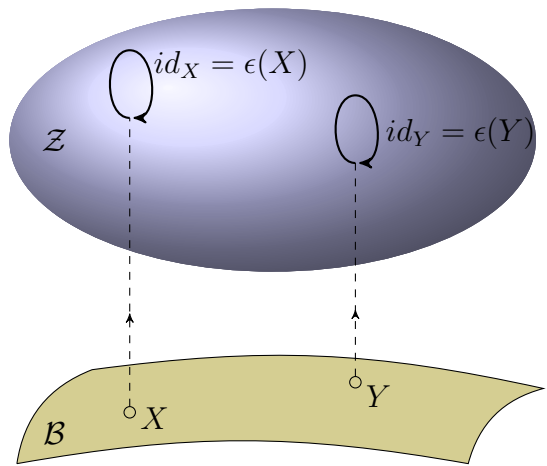

The projection maps, $\alpha$ and $\beta$, map $\mathcal Z$ onto $\mathcal B$. The {\it object inclusion map} $\epsilon$, on the other hand, goes in the opposite direction, from $\mathcal B$ into $\mathcal Z$. It assigns to each object $X$ a very special arrow, called the {\it identity} at $X$, denoted by $id_X=\epsilon(X)$, such that $\alpha(id_X)=\beta(id_X)=X$. Thus, these identities are represented as loop-shaped arrows, as shown in Figure \ref{fig:identities}. 

\begin{figure}[H]
\begin{center}
\begin{tikzpicture} [scale=1.0]
\tikzset{->-/.style={decoration={
  markings,
  mark=at position .5 with {\arrow{stealth'}}},postaction={decorate}}}
  
  \draw[thick, o-o,->-] (3,0) -- (0,0); 
  \draw [thick,o-, ->-] (6,0) -- (3,0);
  \draw[thick,->-](5.85,-0.05) to [bend left] (0.15,-0.05);
  \node[left] at (0,0) {$Z$};
  \node [above] at (3,0) {$Y$};
  \node[right] at (6,0) {$X$};
  \node at (1.5,.7) {$u$};
    \node at (4.5,.7) {$v$};
      \node at (3,-1.3) {$uv$};
\end{tikzpicture}
\end{center}
\caption{Tip-to-tail composition}
\label{fig:compose}
\end{figure}
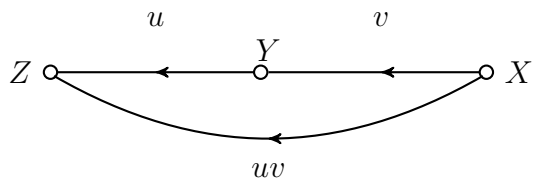

  Given two points, $X$ and $Y$, in the underlying set $\mathcal B$ there may, or may not, be an arrow connecting them. But if there is an arrow $v$ connecting $X$ to $Y$, and another arrow $u$ connecting $Y$ to $Z$, then surely there is the arrow $uv$ connecting $X$ to $Z$. Intuitively, the two original arrows are composed in the natural `end to tail' fashion (concatenation). The notation $uv$ (rather than $vu$) is more suggestive of composition of maps, in the sense that we first apply $v$, as it were, to the point of departure $X$, thus `arriving' at $Y$, and only then we apply $u$ to get to $Z$, as illustrated in Figure \ref{fig:compose}.

\begin{figure}[H]
\begin{center}
\begin{tikzpicture} [scale=1.0]
\tikzset{->-/.style={decoration={
  markings,
  mark=at position .5 with {\arrow{stealth'}}},postaction={decorate}}}
  
  \draw[thick, ->-] (0,0) to [bend left] (6,0); 

  \draw[thick,->-](6,0) to [bend left] (0,0);
  \node[left] at (0,0) {$X$};

  \node[right] at (6,0) {$Y$};
  \node at (3,1.3) {$z$};

      \node at (3,-1.3) {$z^{-1}$};

      \draw[thick, -stealth'] (0,0) arc (360:5:0.5);
          \draw[thick, -stealth'] (6,0) arc (180:-175:0.5);
          \node at (-1.35,0) {$id_X$};
            \node at (7.4,0) {$id_Y$};
\end{tikzpicture}
\end{center}
\caption{Inverse and identities}
\label{fig:inverse}
\end{figure}
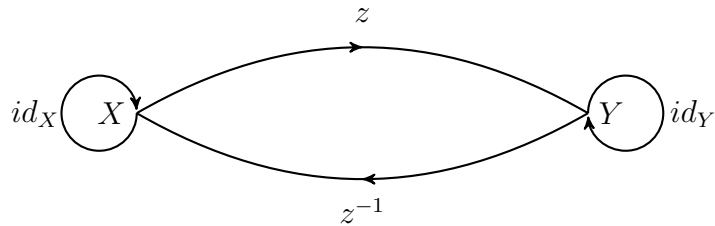

For every arrow $z$ there is necessarily an inverse arrow going in the opposite direction. This feature of a groupoid is clearly reminiscent of the similar idea in the case of a group. It corresponds to the notion of reflexivity of a relation. Finally, since the identities are loop-shaped arrows, they can be composed to the left with all arriving arrows, and to the right with all departing ones, leaving those arrows unchanged. These properties, represented in Figure \ref{fig:inverse}, can be expressed as
\begin{equation} \label{intr4}
z\,\epsilon(\alpha(z))=\epsilon(\beta(z))\,z =z \;\;\;\;\;\;\;\;\;\;zz^{-1}=\epsilon(\beta(z)) \;\;\;\;\;\;\;\;\;z^{-1}z=\epsilon(\alpha(z)).
\end{equation}

\begin{figure}[H]
\begin{center}
\begin{tikzpicture} [scale=1.0]
\tikzset{->-/.style={decoration={
  markings,
  mark=at position .5 with {\arrow{stealth'}}},postaction={decorate}}}
\foreach \x in {1,1.5,...,3}
\draw[thick, ->-] (0,0) arc (270:-90:\x);  
\foreach \x in {1.75,2.25,...,3.25}
\draw[thick, ->-] (0,0) arc (-90:270:\x);  
\draw[fill=black] (0,0) circle(0.1);
\end{tikzpicture}
\end{center}
\caption{A group as a groupoid over a singleton}
\label{fig:group1}
\end{figure}
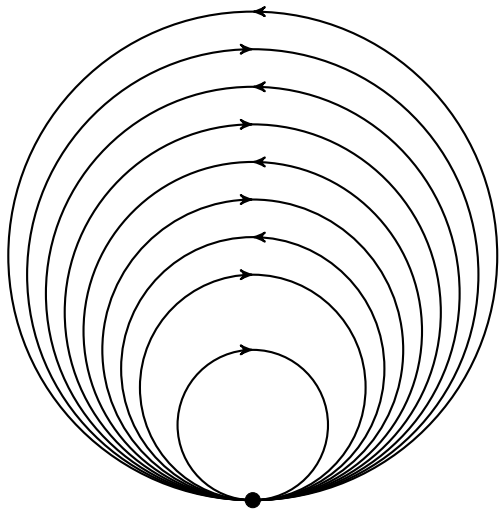

 Not all loop-shaped arrows are units. To understand this point more clearly, we may ask the question: is a group a particular case of a groupoid and, if so, how? If the underlying set $\mathcal B$ happens to be a singleton (a set with just one element), then surely all arrows must be mutually composable! Moreover, there is just a single identity arrow. Thus, a group is a groupoid with a singleton as its base set, as depicted in Figure \ref{fig:group1}.
 
 At this point, we can formulate a complete definition of a groupoid as
 \begin{definition} {\rm A {\it groupoid} ${\mathcal Z}\rightrightarrows{\mathcal B}$ consists of
\begin{enumerate} [(i)]
\item A {\it total set} or {\it set of arrows} $\mathcal Z$,
\item A {\it base set} or {\it set of objects} $\mathcal B$,
\item Two {\it projection maps}, $\alpha: {\mathcal Z} \to {\mathcal B}$ and $\beta: {\mathcal Z} \to {\mathcal B}$, called the {\it source} and the {\it target} maps, respectively,
\item An {\it object inclusion map} $\epsilon: {\mathcal B} \to {\mathcal Z}$, also called the {\it map of identities},
\item A binary operation (called product) in $\mathcal Z$, notated as $uv$ and defined only for those pairs $u,v \in {\mathcal Z}$ such that $\beta(v)=\alpha(u)$, with the following properties:
    \begin{enumerate}[-]
    \item Consistency: $\alpha(uv)=\alpha(v)$ and $\beta(uv)=\beta(u)$, for all $u,v$ such that $\beta(v)=\alpha(u)$,
    \item Associativity: $(uv)w=u(vw)$, whenever the products are defined,
    \item Existence of identities: $z=z\epsilon(\alpha(z))=\epsilon(\beta(z)) z$, for all $z \in {\mathcal Z}$
    \item Existence of inverses: For each $z \in {\mathcal Z}$ there is an inverse arrow $z^{-1} \in {\mathcal Z}$ such that $zz^{-1}=\epsilon(\beta(z))$ and $z^{-1} z = \epsilon(\alpha(z)$.
    \end{enumerate}
\end{enumerate}
}
\end{definition}
 A {\it Lie groupoid} is a groupoid ${\mathcal Z}\rightrightarrows{\mathcal B}$ such that $\mathcal Z$ and $\mathcal B$ are smooth manifolds, all the structure maps are smooth, and the projections are surjective submersions. A thorough treatment of Lie groupoids can be found in \cite{mackenzie}.

\subsection{Examples of groupoids}

\begin{enumerate} [(i)]

\item {\bf Equivalence relation groupoids}: Recall that the {\it Cartesian product} $A \times B$ of two sets, $A$ and $B$, is the collection of all {\it ordered pairs} $(a,b)$, where $a \in A$ and $b \in B$. A {\it relation} $R$ on a set $M$ is any subset of the Cartesian product $M \times M$. An element $m \in M$ is said to be {\it related} to an element $n \in M$ if, and only if, $(m,n) \in R$. An {\it equivalence relation} on $M$ is a relation which is {\it reflexive} ($(m,m) \in R$), {\it symmetric} ($(m,n) \in R \implies (n,m) \in R$), and {\it transitive} ($(m,n) \in R, (n,p) \in R \implies (m,p) \in R$), for all $m,n,p \in R$. It follows from here that we can define a unique groupoid associated with an equivalence relation by identifying: (i) the maps $\alpha$ and $\beta$ with the projections onto the first and second factors of the Cartesian product; (ii) the identity at $m \in M$ with the pair $(m,m)$; and (iii) the inverse of $(m,n)$ with $(n,m)$. The composition can be fitted to the transitivity property (reversing the order of the factors). In the extreme case when $R=M\times M$, we obtain the (transitive, in the terminology of Section \ref{sec:groupgroupoid}) {\it pair groupoid}.  From this point of view, one may say that the concept of groupoid is a generalization of the notion of equivalence relation (since, in a groupoid, two elements may be related in more than one way).
\item {\bf The action groupoid}: A group $ G$ with unit $e$ {\it acts on the left} on a set $M$ if for each $g \in { G}$ there is a map $L_g:M \to M$ such that $L_e(x)=x$ and
$L_g \circ L_h = L_{gh},$ for all $x\in M$, and for all $g,h \in { G}$. A convenient alternative notation for a left action is $gx=L_g(x)$. Thus, the left action properties can be rewritten as
$ex=x$, and $g(hx)=(gh)x$. 
This notation should be used with caution so as not to cause confusion between elements of the group and elements of the set on which the group acts. Given a left action of a group $G$ on a set $M$, we can regard each ordered pair $(x,g)$, where $x \in M$ and $g \in G$, as an arrow with source at $x$ and tip at $gx=L_gx$. If $(x,g)$ and $(y,h)$ are two arrows such that $y=gx$, we can define the composition $(y,g)\circ (x,g)$ as the arrow $(x, hg)$. The identity at $x \in M$ is the arrow $(x,e)$, where $e$ is the group identity. The inverse of the arrow $(x,g)$ is the arrow $(gx,g^{-1})$.
\item {\bf The general linear groupoid $GL({\mathbb R})$}: Considering the collection of all invertible square matrices of all orders as the total set, and the natural numbers $\mathbb N$ as the base set, we can construct a groupoid with matrix multiplication as its partial binary operation. Only matrices of the same order can be multiplied with each other. Thus, this groupoid is precisely the disjoint union of all the general linear groups $GL(n,{\mathbb R})$. In the terminology of Section \ref{sec:groupgroupoid}, these are the vertex groups, and the groupoid is totally intransitive.
\item{\bf The 1-jet groupoid over a manifold $\mathcal B$}: Given a differentiable manifold $\mathcal B$, we denote by $\Pi^1({\mathcal B}, {\mathcal B})$ the set of all linear isomorphisms $L_{X,Y}:T_X{\mathcal B} \to T_Y{\mathcal B}$ between the tangent spaces of all ordered pairs of points $X,Y \in {\mathcal B}$. This set can be regarded as the total set of a groupoid over $\mathcal B$. The composition of maps is identified with the groupoid product operation.
    
\end{enumerate}

\subsection{The groups within the groupoid, local and distant symmetries}
\label{sec:groupgroupoid}

\subsubsection{Transitive and totally intransitive groupoids}

For each object $X$ in $\mathcal B$, we can consider the subset $G_X$ consisting of all loop-shaped arrows that start and end at $X$. More formally,
\begin{equation} \label{intr5}
G_X=\{z\in {\mathcal Z}\;| \; \alpha(z)=\beta(z)=X\}.
\end{equation}
This set is never empty, since it contains at least the identity $id_X$. Moreover, all the arrows in $G_X$ are mutually composable. It follows from a trivial application of the definition of a groupoid that each $G_X$ is a group, called the {\it vertex group} at $X$. The vertex group $G_X$ can be said to represent the local symmetries at $X$.

A groupoid ${\mathcal Z} \rightrightarrows {\mathcal B}$  is {\it transitive} if for every pair of objects $X, Y \in {\mathcal B}$ there is an arrow $z \in {\mathcal Z}$ such that $\alpha(z)=X$ and $\beta(z)=Y$. That is, each pair of objects is connected by at least one arrow.

A groupoid ${\mathcal Z} \rightrightarrows {\mathcal B}$  is {\it totally intransitive}  if for no pair of distinct objects $X \ne Y \in {\mathcal B}$ there is an arrow $z \in {\mathcal Z}$ such that $\alpha(z)=X$ and $\beta(z)=Y$. That is, all arrows of a totally intransitive groupoid are loop-shaped. A totally intransitive groupoid is the union of its vertex groups.

Suppose that two different points, $X$ and $Y$, of $\mathcal B$ are connected by an arrow $z$ (such that $\alpha(z)=X$ and $\beta(z)=Y$). It should be intuitively expected that the local symmetries, embodied in the respective vertex groups $G_X$ and $G_Y$, are in some sense `the same'. Let $g$ belong to $G_X$. We can think of it as a loop-shaped arrow at $X$. But then, according to the tip-to-tail paradigm, the arrow $g'=z g z^{-1}$ is a loop-shaped arrow at $Y$.\footnote{A simple diagram, similar to that used in Figure \ref{fig:inverse}, can be drawn to reinforce the argument.} Since we can reverse the argument by working in the opposite direction, we can convince ourselves that this operation, when applied to every element of either one of the vertex groups, is a bijection. Simply put, there is a one-to-one invertible correspondence between the local symmetries at $X$ and those at $Y$, and this correspondence is of the form  $g'=z g z^{-1}$ for some  fixed $z$. In the parlance of group theory, we may say that if two points of $\mathcal B$ are connected by at least one arrow, the corresponding vertex groups are {\it conjugate}. Consequently, if a groupoid is transitive, all of its vertex groups are mutually conjugate. Choosing any of these groups $G$ as a model, we can say that a transitive groupoid is endowed with a {\it structure group} $G$. This is the closest that a groupoid can get to a group without actually being one.

\subsubsection{Intuitive considerations}

In a classic article \cite{weinstein}, Alan Weinstein explains the intuitive meaning of the groupoid as a conveyor of local and distant symmetries in terms of a tiled floor. Consider an infinite laboratory and a tiling of its infinite floor (${\mathbb R}^2$) with, say, identical hexagons. Each tile is endowed with a number of rotational symmetries, and the floor as a whole is  endowed with a subgroup of the group of translations and rotations of the plane, namely those that exactly match the original pattern. Now suppose that we have a rectangular laboratory of finite extent whose floor is covered by a hexagonal tiling. Clearly, we can no longer speak of any rotational or translational symmetries of the floor as a whole! Intuitively, however, each tile still possesses its own {\it local} rotational symmetries. Clearly, as well, any two distinct tiles are perfectly congruent by some translation (a {\it distant} symmetry). Thus, although we have lost the powerful and elegant mathematical tool afforded by group theory, we still have a feeling that the essential features can be rescued by means of a cognate tool. This tool is precisely the {\it groupoid}.

In the example of the laboratory floor, the vertex groups represent the symmetries possessed by each tile, which we termed local symmetries. To motivate the meaning of all the other arrows of the groupoid, namely those that do not belong to any vertex group or, equivalently, those that are not loop-shaped, let us first consider a floor made from a random arrangement of tiles of different shapes and sizes, somehow patched up together. In this case, there will be no other arrows than those belonging to the vertex groups. Such a groupoid, consisting of the disjoint union of its vertex groups, is called a {\it totally intransitive groupoid}. At the other extreme, consider a floor in which all the tiles have the same shape. We decide to recognize this fact by providing an arrow and its inverse for each pair of tiles. In this case, we call the corresponding groupoid a {\it transitive groupoid}. Each of the non-loop-shaped arrows can be regarded as a {\it distant symmetry}, that is, one arising from comparing the features of two different points, rather than of two states of the same point. If, finally, our proverbial floor has been made by using a few different kinds of tiles, then the groupoid will be neither transitive nor completely intransitive. Some tiles will be mentally connected by arrows and some will not. It is in this sense that a groupoid provides a unifying mathematical structure to encompass both local and distant symmetries.

\section{The material groupoid of an elastic body}
\label{sec:material}

\subsection{Introductory remarks}

Deformable solid bodies are an everyday occurence. In some cases, such as rubber bands and muscles, their deformabiliity is obvious. In other cases, such as the pyramids of Egypt, their deformability can only be determined by precise measurements. Nevertheless, since time immemorial, but certainly since the days of Galileo, Hooke and Newton, at the birth of modern science, it has been universally recognized that the mechanism of load transmission in solid bodies is precisely their ability to deform. Rigid bodies, as useful as they are for understanding the basic laws of mechanics, do not exist.

Confronted with this intellectual challenge and, more dramatically, with the need for designing safe bridges, airplanes, chemical reactors and hip implants, among many other gadgets, mathematicians, physicists, material scientists and engineers of all specialties have produced an enormous body of theoretical, practical and computational knowledge, too large for anyone to fully master or even appreciate. Nevertheless, like in every truly worthy discipline, the fundamental ideas are clearly discernible.

\subsection{Kinematics and constitutive laws}

The first tenet of deformable-body mechanics is that matter is continuous. In mathematical terms, the material body $\mathcal B$ is a differentiable manifold, and thus our discipline can be legitimately encompassed under the umbrella of {\it continuum mechanics}. Now, this continuum assumption is as illusory as the assumption of rigidity. It had been already questioned in antiquity by such luminaries as Democritus and Epicurus, and totally demolished by early twenty-century physics. Nevertheless, even more than one hundred years later, no one can design a garage beam or an airplane wing using quantum mechanics, at least not yet. In rising to a higher (some would say `more phenomenological') level of discourse, continuum mechanics replaces the detailed consideration of the constitution of matter, governed presumably by universal laws, with so-called {\it constitutive laws}, specific to particular materials or particular classes of materials. These `laws' are permeated with parameters to be determined or adjusted by experiments. We often use terms such as {\it material response} and {\it material behaviour} to refer to what is that these laws prescribe: stress, heat flux, free-energy density, and entropy density expressed as some functionals of the temperature and the deformation, a concept that we will presently explain.

A second assumption, related to the first, pertains to the notion of {\it configuration}. The material body $\mathcal B$ manifests itself to us in configurations, which can be conceived as mappings of $\mathcal B$ into our physical space, that is, into the ordinary euclidean space of classical mechanics, where observations and measurements can be made. More importantly, though, these mappings are assumed to be one-to-one (so that no two material points can occupy the same place at the same time) and differentiable as many times as necessary (technically, a configuration is a smooth embedding). A passage from one configuration to another is called a {\it deformation}, which, by composition, inherits all the nice properties of differentiability and invertibility of the configurations. It is true that many phenomena, such as shock waves or mere discontinuities in the material properties, violate the assumed smoothness of the deformations. But these violations are usually confined to surfaces and can be handled with suitable mathematical tools, without affecting the conceptual framework in the bulk of the body.

A third assumption, associated with the so-called {\it elastic materials} is that memory effects are ignored. What is meant by this assumption is that if one knows the state of the material in one configuration (let us call it a {\it reference configuration}), the constitutive laws will depend exclusively on the deformation from this reference configuration. Any other information pertaining to the path while traversing intermediate configurations is irrelevant. Moreover, in a purely mechanical theory, temperature effects and interactions with electromagnetic fields are disregarded. The response depends exclusively on the deformation. Many familiar materials, such as wax, chewing gum, plasticine and other polymers, do not behave this way. So, at best, elasticity is a good approximation for many materials and for some limited periods of time and limited temperature ranges.

Finally, experience with many materials shows that the response is localized. One naive way to picture an elastic deformable body is to imagine a large, but finite, number of point masses interconnected by elastic springs (not necessarily linear in terms of their force-elongation response). In principle, every point may be linked to points near and far, thus giving rise to a so-called {\it non-local} behaviour; but even a limited knowledge of the microscopic nature of crystalline solids leads us to anticipate that it is only the first, or perhaps second, neighbours that matter. In the putative limit as the number of points increases while the density is kept constant, we expect that the material response at a point will depend on the first, or at most the second, derivative of the deformation at that point. For most materials, the first derivative is sufficient. Thus we arrive at the so-called {\it simple elastic materials}.

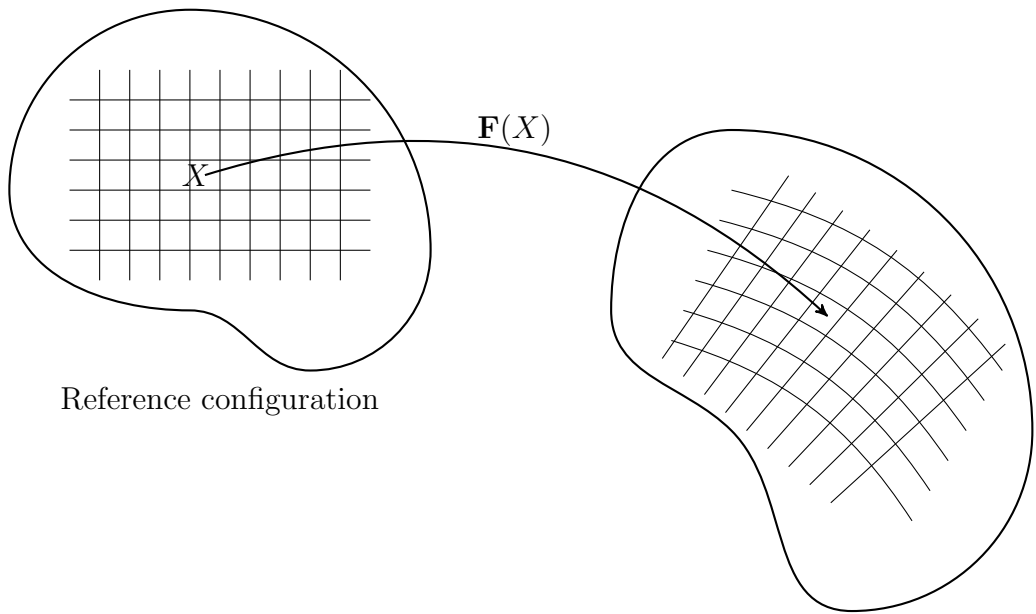
\begin{figure}[H]
\begin{center}
\begin{tikzpicture} [scale=0.8]
\tikzset{->-/.style={decoration={
  markings,
  mark=at position .5 with {\arrow{stealth'}}},postaction={decorate}}}
\draw[thick] (0,0) to [out=90, in=180] (3,3) to [out=0, in=90] (7,-1) to [out=-90, in=0] (5,-3) to [out=180, in=0] (3,-2) to [out=180, in=-90] (0,0);
\draw[thick] (10,-2) to [out=90, in=180] (12,1) to [out=0, in=90] (17,-4) to [out=-90, in=0] (14,-7) to [out=180, in=-45] (12,-4) to [out=135, in=-90] (10,-2);

\foreach \y in {0,1,2,3,4,5}
\draw (1,0.5*\y-1) -- (6,0.5*\y-1);
\foreach \x in {1,2,3,4,5,6,7,8,9}
\draw (0.5*\x+1,-1.5)--(0.5*\x+1,2);
\foreach \y in {0,1,2,3,4,5}
\draw (11+0.2*\y,0.5*\y-2.5) to [bend left=20] (15+0.3*\y,0.5*\y-5.5);
\foreach \x in {1,2,3,4,5,6,7,8,9}
\draw (0.35*\x+10.5,-2.5-0.3*\x)--(0.45*\x+12.5,0.5-0.25*\x-0.01*\x*\x);
\draw[thick,-stealth']  (3.25,0.25) to [bend left] (13.6,-2.1);
\node at (8.4,1) {${\bf F}(X)$};
\node at (3.1,0.25) {$X$};
\node at (3.5,-3.5) {Reference configuration};
\end{tikzpicture}
\end{center}
\caption{The deformation gradient}
\label{fig:potato}
\end{figure}

To eschew in this general introduction the mathematical details, let us think of a body in a reference configuration endowed with a Cartesian coordinate system as a collection of infinitesimal dice aligned with the coordinate axes, as shown in Figure \ref{fig:potato}, where the representation is done for a two-dimensional space for clarity. Under the process of deformation, the body is mapped to another configuration, and the original network of lines is transformed into a curvilinear counterpart. In a first approximation, each die has been transformed into an infinitesimal parallelepiped (a parallelogram in the figure). This is precisely the job of the first derivative of the deformation at each point of the body. Its coordinate representation at each point $X$ is nothing but the Jacobian matrix of the deformation. In the jargon of continuum mechanics, this is a tensor (or linear transformation) known as the {\it deformation gradient}, usually denoted as $\bf F$. According to the explanations above, a constitutive law of a simple elastic body can be expressed as a function $\psi=\psi({\bf F}, X)$, where $\psi$ is a constitutive quantity such as stress or stored energy density.

\subsection{The groupoid representation}

\begin{figure}[H]
\begin{center}
\begin{tikzpicture} [scale=0.8]
\tikzset{->-/.style={decoration={
  markings,
  mark=at position .5 with {\arrow{stealth'}}},postaction={decorate}}}
\draw[thick] (0,0) to [out=90, in=180] (3,3) to [out=0, in=90] (7,-1) to [out=-90, in=0] (5,-3) to [out=180, in=0] (3,-2) to [out=180, in=-90] (0,0);
\draw[thick,o-stealth']  (3.25,0.25) to [bend right] (9,2.5);
\draw[thick,o-stealth']  (4.5,-1) to [bend left] (3.4,0.23);
\node[left] at (3.6,-0.5) {${\bf P}_{12}$};
\node at (8.4,1) {${\bf F}$};
\node at (8.4,0.5) {arbitrary};
\node[above] at (3.1,0.25) {$X_2$};
\node[below] at (4.5,-1) {$X_1$};
\node at (3.5,-3.5) {Reference configuration};
\end{tikzpicture}
\end{center}
\caption{A material isomorphism as a transplant}
\label{fig:transplant}
\end{figure}
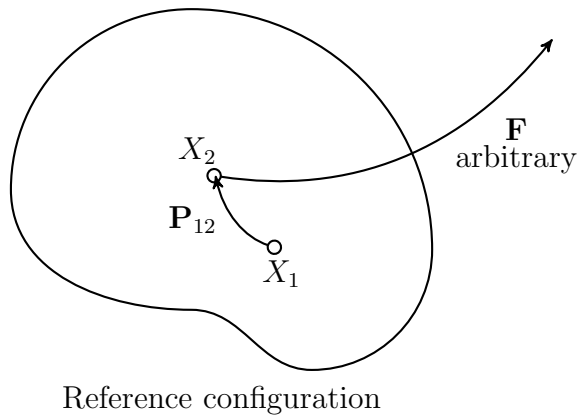

Inspired by the example of the tiled floor, we ask the question: given two different points, $X_1$ and $X_2$, in the body, are they made of the same material? We say that two points are {\it materially isomorphic} if there is a linear map ${\bf P}_{12}$ between the die at $X_1$ and the die at $X_2$ such that the equation
\begin{equation} \label{intr6}
\psi({\bf FP}_{12}, X_1)=\psi({\bf F},X_2),
\end{equation}
is satisfied identically for all possible deformation gradients $\bf F$, as suggested in Figure \ref{fig:transplant}. Physically, using a somewhat forced surgical analogy, what this means is that we have cut a piece of skin of a patient and we have stretched or otherwise distorted it before transplanting it to a damaged area in another part of the body. This transplant is precisely ${\bf P}_{12}$. The transplant is successful (namely, the two points are materially isomorphic) if, and only if, we can no longer distinguish between the original response of the tissue before it was damaged and the response of the implant after it had been deformed by ${\bf P}_{12}$. If we assign to each material isomorphism between a pair of body points an arrow, we obtain a groupoid ${\mathcal Z} \rightrightarrows {\mathcal B}$ called the {\it material groupoid} induced by the given constitutive law.

The vertex groups of the material groupoid are precisely the local symmetry groups of the material, namely, linear transformations that leave the response invariant at a point. If $G_1$ is the vertex group at $X_1$, and ${\bf P}_{12}$ is a material isomorphism from $X_1$ to $X_2$, then $g{\bf P}_{12}$ is also a material isomorphism between these two points. Thus, the collection $\mathcal P$ of the material groupoid arrows between these points can be expressed as
\begin{equation} \label{eq:arrows}
{\mathcal P}={\bf P}_{12} G_1.
\end{equation}

\subsection{Uniformity and homogeneity}

 A material is said to be uniform if it is made of the same material at all points. It follows that a body is {\it uniform} if, and only if, its material groupoid is transitive. Indeed, transitivity means that all pairs of points are connected by arrows. The body is smoothly uniform if its material groupoid is a transitive Lie groupoid. In other words, every point has a neighbourhood in which the material isomorphisms can be chosen smoothly.

A subtler question pertains to the notion of {\it homogeneity}. A smoothly uniform body is (locally) homogeneous if there is a deformation that can bring it to a configuration in which mere translations (within the neighbourhood) are material isomorphisms. A body may be smoothly uniform, but we may not be able to achieve this goal. In this case, we say that the body is {\it inhomogeneous}. In other words, we are not able to bring all the points (in a neighbourhood) simultaneously to an identical state. A common example is the presence of {\it residual stresses} in biological materials, such as arteries. Another example is furnished by metals that have been loaded beyond their elastic range and then unloaded, as we can demonstrate by significantly twisting a metal paper clip and then placing it, unloaded, on a table.

 Homogeneity is a question of {\it integrability}, a question whose precise answer requires considerable additional mathematical apparatus. In the simplest scenario, when the symmetry group of a smoothly uniform body is discrete (such as is the case with crystal classes), the smooth material isomorphisms (in a neighbourhood) give rise to a unique {\it distant parallelism}. Two vecotrs at two distinct points are materially parallel if they are mapped into each other by a material isomorphism. In differential geometric terms, we have a connection with vanishing curvature and possibly non-vanishing torsion. The condition of (local) homogeneity is precisely the vanishing of the torsion of this unique material connection.

In the more general case, in which the symmetry group of the material is a Lie group (such as is the case for isotropic or transversely isotropic solids), the homogeneity question can be formulated in terms of the material groupoid or the corresponding material algebroid. Material groupoids, as defined, are subgroupoids of the 1-jet groupoid of ${\mathbb R}^3$. This circumstance affords a notion of integrability akin to the concept of homogeneity. A material groupoid is integrable if it is `equivalent' (locally diffeomorphic) to a reduction of the 1-jet groupoid of ${\mathbb R}^3$. Within this conceptual framework, local homogeneity is equivalent to integrability of the material groupoid. These issues are treated in great detail in \cite{victor1}.

\subsection{Binary composites}
\label{sec:binary}

A composite is a material consisting of a permanent blend of two or more materials. Each of the components is still identifiable in the mix, but the composite behaves as a single body $\mathcal B$. In particular, the deformation and its gradient are the same for all the components. In the binary case, consider two elastic materials, $\mathcal M$ and $\hat{\mathcal M}$, each of which, when regarded independently, makes up a smoothly uniform body. The respective material groupoids, ${\mathcal Y} \rightrightarrows {\mathcal B}$ and ${\hat{\mathcal Y}}\rightrightarrows {\mathcal B}$, therefore, are transitive Lie groupoids, with respective structure groups $\mathcal G$ and $\hat{\mathcal G}$.

Assuming that the bonding of the materials has not involved any chemical reactions, it is reasonable to assume that the mechanical response of the composite will be the result of a weighted combination of the responses of the component materials. The material groupoid ${\mathcal Z} \rightrightarrows {\mathcal B}$ will accordingly consist of the intersection ${\mathcal Z}={\mathcal Y}\cap{\hat{\mathcal Y}}$ of the material groupoids of the components.\footnote{For Lie groupoids, the intersection of the manifolds involved may need to satisfy an extra condition, such as that of {\it clean intersection} \cite{bott}.} In simpler terms, the material groupoid of the composite will consist of all the arrows that are common to the material groupoids of the components. Consequently, each vertex groupoid of the intersection is precisely the intersection of the vertex groupoids of the components.

As far as the remaining common arrows are concerned, there are several possibilities. At one extreme, it is possible that, beyond the vertex groups, there are no common arrows. In this case, although the components are transitive, the material groupoid of the composite is totally intransitive. At the other extreme, if there is at least one common arrow between each pair of points, the material groupoid of the composite is transitive.

 From the physical standpoint, the transitive case corresponds to the fact that the two components have been blended consistently in the same way at all body points. We will, therefore, declare that {\it a binary composite is uniform if the intersection of the material groupoids of its components is a transitive groupoid}. Intermediate cases between transitivity and total intransitivity can be interpreted as functionally graded materials, or perhaps laminates and other partially uniform arrangements.

A similar analysis can be carried out with respect to homogeneity. If the two solid components are homogeneous, and if they share a homogeneous configuration, the composite is uniform and homogeneous, but there may not be a homogeneous configuration that is also stress-free. If the two homogeneous components do not share any homogeneous configuration, the composite cannot be uniform, let alone homogeneous.

\begin{example}  {\rm For the purpose of illustration it is convenient to work in a two-dimensional setting. If the symmetry group of a point is the orthogonal group, this will be indicated with a circle. If it is a conjugate of the orthogonal group it will be represented as an ellipse. A discrete symmetry group, and most particularly the trivial symmetry group, will be indicated with the symbol $\perp$. Further clarification of this rather imprecise representation will be provided in each example.
\begin{enumerate}
\item {\bf Uniformity and homogeneity preserved}: Consider two materially uniform, homogeneous and stres-free plates as shown in Figure \ref{fig:item1}. The material isomorphisms in the first plate are rotations, while in the second plate they are identities. Translations are trivially factored out from any material isomorphism.
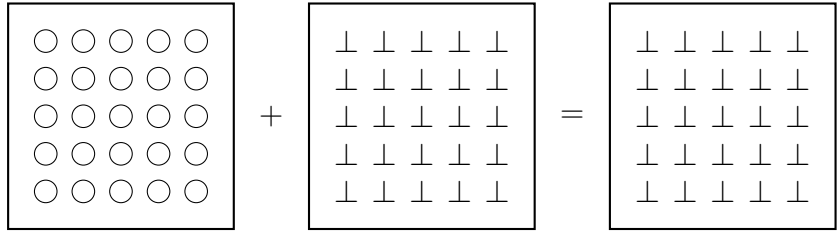
\begin{figure}[H]
\begin{center}
\begin{tikzpicture}[scale=0.5]
\draw[thick] (0,0)--(6,0)--(6,6)--(0,6)--cycle;
\foreach \x in {1,2,3,4,5}
\foreach \y in {1,2,3,4,5}
\draw (\x,\y) circle (0.3);
\begin{scope}[xshift=8cm]
\draw[thick] (0,0)--(6,0)--(6,6)--(0,6)--cycle;
\foreach \x in {1,2,3,4,5}
\foreach \y in {1,2,3,4,5}
\node at (\x,\y) {$\perp$};
\end{scope}
\begin{scope}[xshift=16cm]
\draw[thick] (0,0)--(6,0)--(6,6)--(0,6)--cycle;
\foreach \x in {1,2,3,4,5}
\foreach \y in {1,2,3,4,5}
\node at (\x,\y) {$\perp$};
\end{scope}
\node at (7,3) {$+$};
\node at (15,3) {$=$};
\end{tikzpicture}
\end{center}
\caption{Two uniform and homogeneous plates}
\label{fig:item1}
\end{figure}

\item {\bf Uniformity preserved}: If, in the previous example, the second plate is replaced with a uniform, but inhomogeneous, plate, the result is a uniform, but inhomogeneous, composite (as suggested in Figure \ref{fig:item2}). The material isomorphisms of the second plate are pure rotations so that, if the points are in a natural stress-free state, so will the composite be (in a state known as curvilinear or contorted aelotropy).
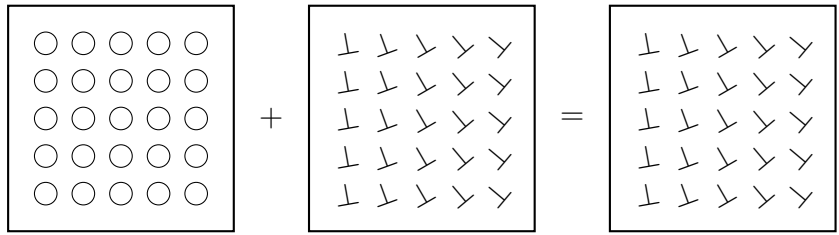
\begin{figure}[H]
\begin{center}
\begin{tikzpicture}[scale=0.5]
\draw[thick] (0,0)--(6,0)--(6,6)--(0,6)--cycle;
\foreach \x in {1,2,3,4,5}
\foreach \y in {1,2,3,4,5}
\draw (\x,\y) circle (0.3);
\begin{scope}[xshift=8cm]
\draw[thick] (0,0)--(6,0)--(6,6)--(0,6)--cycle;
\foreach \x in {10,20,30,40,50}
\foreach \y in {1,2,3,4,5}
\node at (0.1*\x,\y) {\rotatebox[origin=c]{\x}{$\perp$}};
\end{scope}
\begin{scope}[xshift=16cm]

\draw[thick] (0,0)--(6,0)--(6,6)--(0,6)--cycle;
\foreach \x in {10,20,30,40,50}
\foreach \y in {1,2,3,4,5}
\node at (0.1*\x,\y) {\rotatebox[origin=c]{\x}{$\perp$}};
\end{scope}
\node at (7,3) {$+$};
\node at (15,3) {$=$};
\end{tikzpicture}
\end{center}
\caption{Two uniform  plates}
\label{fig:item2}
\end{figure}

\item {\bf Loss of uniformity}: If the first plate is also uniform but not homogeneous, the uniformity will be lost in general. In the case of Figure \ref{fig:item3}, both plates are of the same kind, but the rotation fields involved in their respective material isomorphisms do not coincide. As a result, the composite is non-uniform, such as can be the case in functionally graded bodies.
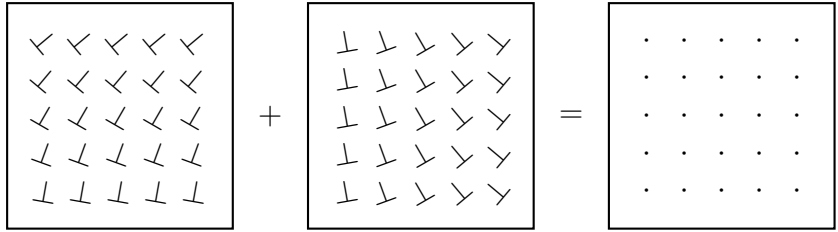
\begin{figure}[H]
\begin{center}
\begin{tikzpicture}[scale=0.5]
\draw[thick] (0,0)--(6,0)--(6,6)--(0,6)--cycle;
\foreach \x in {1,2,3,4,5}
\foreach \y in {10,20,30,40,50}
\node at (\x, 0.1*\y) {\rotatebox[origin=c]{-\y}{$\perp$}};
\begin{scope}[xshift=8cm]
\draw[thick] (0,0)--(6,0)--(6,6)--(0,6)--cycle;
\foreach \x in {10,20,30,40,50}
\foreach \y in {1,2,3,4,5}
\node at (0.1*\x,\y) {\rotatebox[origin=c]{\x}{$\perp$}};
\end{scope}
\begin{scope}[xshift=16cm]

\draw[thick] (0,0)--(6,0)--(6,6)--(0,6)--cycle;
\foreach \x in {10,20,30,40,50}
\foreach \y in {1,2,3,4,5}
\node at (0.1*\x,\y) {$\cdot$};
\end{scope}
\node at (7,3) {$+$};
\node at (15,3) {$=$};
\end{tikzpicture}
\end{center}
\caption{Loss of uniformity from two uniform  plates}
\label{fig:item3}
\end{figure}

\item {\bf Loss of stress-free configurations}: In Figure \ref{fig:item4}, ellipses indicate conjugates of the orthogonal group. Physically, this can be interpreted as having stretched the plate before attaching it to the seond  plate. Both plates are uniform and homogeneous, and so is the resulting composite, since it possesses a configuration in which all the material isomorphisms are Euclidean translations. Nevertheless, the resulting composite does not admit a stress-free configuration. This feature is absent in the case of a single elastic homogeneous solid, which can always be brought trivially to a stress-free configuration.
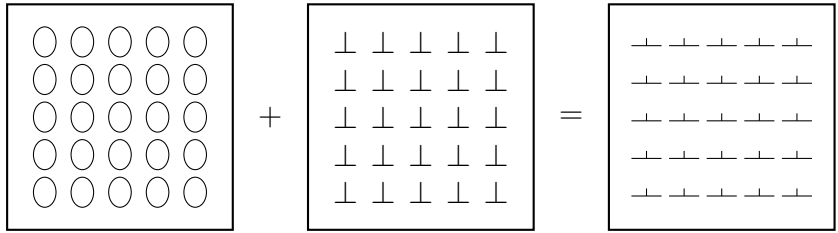
\begin{figure}[H]
\begin{center}
\begin{tikzpicture}[scale=0.5]
\draw[thick] (0,0)--(6,0)--(6,6)--(0,6)--cycle;
\foreach \x in {1,2,3,4,5}
\foreach \y in {1,2,3,4,5}
\draw (\x,\y) ellipse (0.3 and 0.4);
\begin{scope}[xshift=8cm]
\draw[thick] (0,0)--(6,0)--(6,6)--(0,6)--cycle;
\foreach \x in {1,2,3,4,5}
\foreach \y in {1,2,3,4,5}
\node at (\x,\y) {$\perp$};
\end{scope}
\begin{scope}[xshift=16cm]
\draw[thick] (0,0)--(6,0)--(6,6)--(0,6)--cycle;
\foreach \x in {1,2,3,4,5}
\foreach \y in {1,2,3,4,5}
{\draw (\x-0.4,\y-0.1)--(\x+0.4,\y-0.1);
\draw (\x,\y-0.1)--(\x,\y+0.1);}
\end{scope}
\node at (7,3) {$+$};
\node at (15,3) {$=$};
\end{tikzpicture}
\end{center}
\caption{Loss of stress-free configurations from two homogeneous plates}
\label{fig:item4}
\end{figure}
\end{enumerate}
}
\end{example}

The analysis presented above is by no means exhaustive, but it serves to demonstrate how the properties of uniformity and homogeneity of a composite may be derived from their counterparts at the component level. A question that arises naturally is whether there is a combined mathematical structure characterizing the composite and incorporating the material groupoids of the components into a single overarching geometric structure. The positive answer is provided by double groupoids, which have been successfully applied to other physical areas, such as quantum field theory \cite{kerler}. Although the presentation offered in the next section is limited to the case of binary composites, it is to be expected that the formulation may be extended to more general mixtures, and also to non-local theories, in which the material response of a body is expressed in terms of functions involving pairs of body points.

\section{Double groupoids}
\label{sec:doublegroupoid}

\subsection{Definition and terminology}

A {\it double groupoid} consists of a set $\mathcal Q$ endowed with two groupoid structures, ${\mathcal Q} \rightrightarrows {\mathcal V}$ and ${\mathcal Q} \rightrightarrows {\mathcal H}$, called, respectively, the {\it horizontal} and the {\it vertical} structures. Each of the base spaces is itself a groupoid over a common base $\mathcal B$, namely ${\mathcal H}\rightrightarrows {\mathcal B}$ and ${\mathcal V}\rightrightarrows {\mathcal B}$, called, respectively, the horizontal and vertical side groupoids. A common notation for a double groupoid is
\begin{equation} \label{dg1}
\begin{matrix}
{\mathcal Q}                   &\rightrightarrows  & {\mathcal V} \\
  \downdownarrows    &                      &  \downdownarrows  \\
  {\mathcal H}                 & \rightrightarrows   &\mathcal{B}
\end{matrix}
\end{equation}
Since there are 4 different groupoids involved in the definition, it is necessary to establish a notation for the elements, maps, and operations of each. Authors differ in this respect, each author adopting a notation that seems more convenient to particular applications. The notations adopted herein are in part a mixture of the notations in \cite{ehresmann}, \cite{brown1}, and \cite{natale}. 

The elements of $\mathcal B$ will be denoted with capital letters $(X, Y, ...)$. The arrows, units, and projections of the side groupoid ${\mathcal H}\rightrightarrows {\mathcal B}$  will follow the usual notation for groupoids used in Section \ref{sec:groupoids}. For the vertical side groupoid  ${\mathcal V}\rightrightarrows {\mathcal B}$, the same notation will be used, except that a hat (circumflex accent) will be added. The product will be indicated by simple apposition, since there is no room for confusion. Thus, for example, $uv$ is an arrow in $\mathcal H$, and ${\hat p} {\hat q}$ is an arrow in $\mathcal V$. The projections and units will be also distinguished by the presence or absence of a hat, to avoid any possible ambiguity. Thus, $\alpha(u)$ is the source of $u \in {\mathcal H}$ and ${\hat \alpha}({\hat p})$ is the source of ${\hat p} \in {\mathcal V}$ .

The elements of $\mathcal Q$ will be denoted with double-strike letters $({\mathbb A},{\mathbb B}, ...)$, and represented graphically as squares (much in the same way as the elements of $\mathcal H$ and $\mathcal V$ are represented as arrows). A square has two horizontal parallel sides belonging to $\mathcal H$, while the vertical sides belong to $\mathcal V$. Moreover, there are two different products between squares, one for each of the two groupoid structures  ${\mathcal Q}\rightrightarrows {\mathcal V}$ and ${\mathcal Q}\rightrightarrows {\mathcal H}$, denoted respectively by $\hor$ and $\ver$.

The projections and the units in each of the two bundle structures of $\mathcal Q$ will be signaled with a superimposed tilde and with a subscript $_H$ or $_V$ for the horizontal and vertical structures, respectively. Note that the source and target of the horizontal structure belong to $\mathcal V$, while their couterparts for the vertical structure belong to $\mathcal H$. To exemplify this notation, we draw an element ${\mathbb A} \in {\mathcal Q}$  in the shape of a square, as anticipated, namely,
\begin{equation} \label{dg2}
\begin{tikzpicture}[baseline=(current  bounding  box.center)]
  \draw[thick,-stealth'] (3,0) -- (0,0);
  \draw[thick,-stealth'] (3,3) -- (0,3);
  \draw[thick,-stealth'] (3,0) -- (3,3);
  \draw[thick,-stealth'] (0,0) -- (0,3);
\node at (1.5,1.5) {$\mathbb A$};
\node at (4.2,1.5) {${\hat s}={\tilde \alpha}_H({\mathbb A})$};
\node at (-1.3,1.5) {${\hat t}={\tilde \beta}_H({\mathbb A})$};
\node[above] at (1.5,3) {$t={\tilde \beta}_V({\mathbb A})$};
\node[below] at (1.5,0) {$s={\tilde \alpha}_V({\mathbb A})$};
\end{tikzpicture}
\end{equation}

In the horizontal structure, the product ${\mathbb A}\hor{\mathbb A}'$ can only be carried out if ${\hat s}={\hat t}'$, in agreement with the usual concatenation criterion. Graphically,
\begin{equation} \label{dg3}
\begin{tikzpicture}[baseline=(current  bounding  box.center)]
\foreach \x in {0,1.5,4.5}
{  \draw[thick,-stealth'] (1.5+\x,0) -- (0+\x,0);
  \draw[thick,-stealth'] (1.5+\x,1.5) -- (0+\x,1.5);
  \draw[thick,-stealth'] (1.5+\x,0) -- (1.5+\x,1.5);
  \draw[thick,-stealth'] (0+\x,0) -- (0+\x,1.5);}
\node[left] at(-0.,0.75) {${\mathbb A}\hor {\mathbb A}'\;=\;{\hat t}$};
\node[above] at(0.75,1.5) {$  t$};
\node[above] at(2.25,1.5) {$ t'$};
\node[below] at(0.75,-0.1) {$  s$};
\node[below] at(2.25,0) {$ s'$};
\node[right] at (3,0.75) {${\hat s}'\;=\;{\hat t}$};
\node[above] at (5.25,1.5) {$tt'$};
\node[below] at (5.25,0) {$ss'$};
\node[right] at (6,0.75) {${\hat s}'$};
\end{tikzpicture}
\end{equation}
Similarly, in the vertical structure, the product ${\mathbb A}\ver{\mathbb A}'$ can only be carried out if $s'=t'$, or graphically
\begin{equation} \label{dg4}
\begin{tikzpicture}[baseline=(current  bounding  box.center)]
\foreach \x in {0,-1.5}
{  \draw[thick,-stealth'] (1.5,0+\x) -- (0,0+\x);
  \draw[thick,-stealth'] (1.5,1.5+\x) -- (0,1.5+\x);
  \draw[thick,-stealth'] (1.5,0+\x) -- (1.5,1.5+\x);
  \draw[thick,-stealth'] (0,0+\x) -- (0,1.5+\x);}
\node[left] at(-0.,0) {${\mathbb A}\ver {\mathbb A}'\;=\;\;\;\;\;$};
\node[left] at (0,0.75) {${\hat t}$};
\node[above] at(0.75,1.5) {$  t$};
\node[right] at (1.5,0.75) {${\hat s}$};
\node[right] at (1.5,-0.75) {${\hat s}'$};
\node[below] at(0.75,-1.5) {$s'$};
\node[left] at (0,-0.75) {$\hat t'$};
\node[right] at (1.8,0.1) {$\;\;=\;\;\;{\hat t} {\hat t}'$};
\foreach \x in {3.5}
{\draw[thick,-stealth'] (1.5+\x,-0.75) -- (0+\x,0-0.75);
  \draw[thick,-stealth'] (1.5+\x,1.5-0.75) -- (0+\x,1.5-0.75);
  \draw[thick,-stealth'] (1.5+\x,0-0.75) -- (1.5+\x,1.5-0.75);
  \draw[thick,-stealth'] (0+\x,0-0.75) -- (0+\x,1.5-0.75);}
\node[right] at (5,0.1) {${\hat s}{\hat s}'$};
\node[above] at (4.25,0.75) {$t$};
\node[below] at (4.25,-0.75) {$s'$};
\end{tikzpicture}
\end{equation}

The definition of a double groupoid must be completed by the imposition of a compatibility condition between the two structures, namely,
\begin{equation} \label{dg5}
({\mathbb A}\hor{\mathbb B})\ver({\mathbb C}\hor{\mathbb D})=({\mathbb A}\ver{\mathbb C})\hor({\mathbb B}\ver{\mathbb D}),
\end{equation}
whenever the operations are possible. Graphically, this condition means that, in a square made of four squares whose edges in contact match, the result is the same whether one first composes horizontally and then vertically, or vice versa. In other words, the large square shown below makes sense.
\begin{equation} \label{dg6}
\begin{tikzpicture}[baseline=(current  bounding  box.center)]
\foreach \x in {0,1.5}
{\foreach \y in {0,1.5}
{  \draw[thick,-stealth'] (1.5+\x,0+\y) -- (0+\x,0+\y);
  \draw[thick,-stealth'] (1.5+\x,1.5+\y) -- (0+\x,1.5+\y);
  \draw[thick,-stealth'] (1.5+\x,0+\y) -- (1.5+\x,1.5+\y);
  \draw[thick,-stealth'] (0+\x,0+\y) -- (0+\x,1.5+\y);}}
\node at (0.75,0.75) {$\mathbb C$};
\node at (0.75,2.25) {$\mathbb A$};
\node at (2.25,0.75) {$\mathbb D$};
\node at (2.25,2.25) {$\mathbb B$};
\end{tikzpicture}
\end{equation}

Finally, the {\it double source map} assigning to each square $\mathbb A$ its sources in the horizontal and vertical structures, ${\tilde \alpha}_H({\mathbb A}), {\tilde \alpha}_V({\mathbb A})$, is often assumed to be surjective. The graphical representation of this assumption is that a right angle made of a horizontal and a vertical arrow issuing from a common corner in $\mathcal B$, can always be completed (not necessarily uniquely) to a square in $\mathcal Q$, as suggested below. This is known as the {\it filling condition}.
\begin{equation} \label{dg7}
\begin{tikzpicture}[baseline=(current  bounding  box.center)]
\draw[thick,-stealth'] (0,0)--(-2,0);
\draw[thick,-stealth'] (0,0)--(0,2);
\draw[thick,-stealth'] (4,0)--(2,0);
\draw[thick,-stealth'] (4,0)--(4,2);
\draw[thick,-stealth'] (4,2)--(2,2);
\draw[thick,-stealth'] (2,0)--(2,2);
\node at (1,1) {$\Longrightarrow$};
\end{tikzpicture}
\end{equation}

A useful example of a double groupoid is the {\it coarse double groupoid}, $\square(H,V)$, generated by two groupoids with a common base set. It consists of all the possible consistent squares that can be formed using the two groupoids as sides. Any double groupoid can be naturally mapped as an inclusion into the coarse double groupoid generated by its side groupoids. A {\it double Lie groupoid} is obtained when the four groupoids involved in the algebraic definition of a double groupoid are Lie groupoids, with the additional condition that the double source map is a surjective submersion.

\subsection{Transitivity}

\subsubsection{Definitions}

Recalling that a groupoid is transitive if every pair of objects is connected by at least one arrow, the transitivity of the horizontal structure ${\mathcal Q} \rightrightarrows {\mathcal V}$ can be interpreted as follows: every pair of vertical arrows can be completed to yield at least one square in $\mathcal Q$. A similar statement can be applied, mutatis mutandis, to the vertical structure ${\mathcal Q} \rightrightarrows {\mathcal H}$. A possible definition of the concept of transitivity for a double groupoid is, therefore, the following: A double groupoid is said to be transitive if all 4 constituent grupoids are transitive. Put differently, for each pair of objects in $\mathcal B$ there are at least one horizontal arrow and one vertical arrow that join them, and for each pair of vertical (horizontal) arrows, there is at least one horizontal (vertical) pair with which to complete a square.

In an important article \cite{brown1} on double groupoids, however, Brown and Mackenzie propose a somewhat stronger definition of transitivity, which they refer to as {\it local triviality}. In its simplified form, this definition requires that for every three-sided arrow arrangement there is at least one arrow that closes it into a square in $\mathcal Q$. In other words, every configuration of matching sides in the shape of  $\sqcup$ or $\sqsupset$ can be completed to a square in $\mathcal Q$. A double groupoid that is transitive in the usual sense may fail to be transitive  (locally trivial) under this more stringent requirement for the transitivity of the higher structures.

\subsubsection{The core groupoid}

Consider the subset ${\mathcal K} \subset {\mathcal Q}$ of all squares whose horizontal and vertical sources are the vertical and horizontal units of a point $X \in {\mathcal B}$, that is, all squares of the form
\begin{equation} \label{dg8}
\begin{tikzpicture}[baseline=(current  bounding  box.center)]
\draw[thick,-stealth'] (4,0)--(2,0);
\draw[thick,-stealth'] (4,0)--(4,2);
\draw[thick,-stealth'] (4,2)--(2,2);
\draw[thick,-stealth'] (2,0)--(2,2);
\node[above] at (3,2) {$t$};
\node[left] at (2,1) {$\hat t$};
\node[right] at (4,1) {${\hat s}={\hat \epsilon}(X)$};
\node[below] at (3,0) {$s=\epsilon(X)$};
\node at (3,1) {$\mathbb K$};
\end{tikzpicture}
\end{equation}
as $X$ ranges over $\mathcal B$. In the elegant notation used in \cite{natale}, these elements are represented as
\begin{equation} \label{dg9}
\begin{tikzpicture}[baseline=(current  bounding  box.center)]
\draw[thick] (0,0)--(0,2)--(2,2);
\draw[thick, double, double distance=1pt] (0,0)--(2,0)--(2,2);
\end{tikzpicture}
\end{equation}
where the identities are indicated by double lines.

The set $\mathcal K$ is endowed with a canonical groupoid structure ${\mathcal K}\rightrightarrows{\mathcal B}$, called the {\it core groupoid} of $\mathcal Q$. Instead of providing a formal definition of the structure maps and operations in the core groupoid, it is sufficient to observe that the elements of $\mathcal K$ can be regarded as pairs of arrows $(t,{\hat t})$ (one each from $\mathcal H$ and $\mathcal V$) with a common source and a common target in $\mathcal B$. Multiplication, inversion and units are obtained by independently using the appropriate counterparts in each of the side groupoids. To show that this simple rule can be achieved via a judicious application of the operations in the double groupoid $\mathcal Q$ of departure, it is sufficient to verify the multiplication in $\mathcal K$ graphically in the following diagram
\begin{equation} \label{dg10}
\begin{tikzpicture}[baseline=(current  bounding  box.center)]
\node at(-2,-0.25) {$(t,{\hat t})(t',{\hat t}')\;=\;$};
\draw[thick] (0,0)--(0,2)--(2,2);
\node[left] at (0,1) {$\hat t$};
\node[above] at (1,2) {$t$};
\draw[thick, double, double distance=1pt] (0,0)--(2,0)--(2,2);
\draw[thick] (2.5,2)--(4.5,2);  
\node[above] at (3.5,0) {$t'$};
\node[above] at (3.5,2) {$t'$};
\node[below] at (3.5,-0.5) {$t'$};
\node[left] at (2.5,-1.5) {${\hat t}'$};
\draw[thick] (2.5,0)--(4.5,0);
\draw[thick, double, double distance=1pt](2.5,2)--(2.5,0);
\draw[thick, double, double distance=1pt](4.5,2)--(4.5,0);
\draw[thick] (2.5,-2.5)--(2.5,-0.5)--(4.5,-0.5);
\draw[thick, double, double distance=1pt] (2.5,-2.5)--(4.5,-2.5)--(4.5,-0.5);
\node at(6,-0.25) {$\;=\;(tt',{\hat t}{\hat t}')$};
\end{tikzpicture}
\end{equation}
The intermediary is nothing but a unit in the horizontal structure of $\mathcal Q$. the same result is obtained invoking a unit in the vertical structure.

In a more intuitive picture, one may say that the arrows of the core groupoid are, essentially, pairs of arrows (one horizontal and one vertical) between the same sources and targets, as one can gather from the diagrams of Equations (\ref{dg8}) or (\ref{dg9}). From this observation, it follows that the core groupoid is endowed with two canonical maps, $\partial_H: {\mathcal K} \to {\mathcal H}$, and $\partial_V:{\mathcal K} \to {\mathcal V}$, given, respectively, by $\partial_H{\mathbb K}=t={\tilde \beta}_V({\mathbb K})$ and $\partial_V{\mathbb K}={\hat t}={\tilde \beta}_H({\mathbb K})$. The triple $\{{\mathcal K}, \partial_H, \partial_V\}$ is the {\it core diagram} of the double groupoid. A crucial theorem \cite{brown1} asserts that the core groupoid of a locally trivial double groupoid determines the double groupoid up to isomorphism. Andruskiewitsch and Natale \cite{natale} obtain a more general result for algebraic double groupoids in terms of their {\it frame}, which can be loosely defined as the image of the insertion map of the double groupoid into the coarse double groupoid generated by its side groupoids.

\subsection{Application to binary composites}

\subsubsection{The material double groupoid}
\label{sec:materialdg}

\begin{definition} The material double groupoid of a binary composite is the double groupoid with base $\mathcal B$, whose side groupoids are the material groupoids of the components, and whose squares $\mathbb Q$ satisfy the condition
\begin{equation} \label{dg11}
   {\tilde \beta}_V({\mathbb Q})\;\;{\tilde \alpha}_H({\mathbb Q})={\tilde \beta}_H({\mathbb Q})\;\; {\tilde \alpha}_V({\mathbb Q}).
\end{equation}
\end{definition}

In the notation of the diagram in Equation (\ref{dg2}), the condition of the material double groupoid reads
\begin{equation} \label{dg12}
t{\hat s} ={\hat t}s.
\end{equation}
In words, we define the material double groupoid as the collection of commuting squares made up from arrows of the two side groupoids. For this definition to make sense, it is necessary that the mixed composition of arrows from the side groupoids be defined. In the case of the material double groupoid, this is automatically the case, since both material groupoids are subgroupoids of the 1-jet groupoid of ${\mathbb R}^3$.

\begin{prop}  A binary composite is uniform if, and only if, the core of its material double groupoid is a transitive groupoid with structure group equal to the intersection of the structure groups of the (transitive) side groupoids.
\end{prop}
\begin{proof} If the core of the material double groupoid is transitive, for any pair of points, $X,Y \in {\mathcal B}$, there is a square ${\mathbb K}$, such as the one shown in the diagram of Equation (\ref{dg8}), with $\beta(t)={\hat \beta}({\hat t})=Y$. Applying Equation (\ref{dg12}), we obtain $t={\hat t}$. It follows that both side groupoids are transitive and that they have at least one arrow in common between any pair of points. Therefore, according to the criterion of Section \ref{sec:binary}, the composite is uniform. Moreover, the collection of all the common arrows between $X$ and $Y$ is generated by the intersection $G_X \cap{\hat G}_X$ acting on the left on $t$ and $\hat t$. The proof of the converse is trivial.
\end{proof}

The condition of local triviality of the material double groupoid is much stronger than the mere transitivity of the core groupoid. Indeed, if the double groupoid is locally trivial, we can choose three-sided channels (of the shapes $\sqcup$ or $\sqsupset$) and make two adjacent sides equal to unit arrows, thus recovering the core. For {\it every} choice of the third side, however, according to the definition of local triviality, there must be a corresponding closing side. This fact (combined with the defining condition of the material double groupoid) effectively means that the vertex groups of the side groupoids are pointwise identical. Thus, a composite with a locally trivial material double groupoid has the same uniformity and symmetry as its components.

\begin{example} {\rm {\bf A locally trivial composite}:  Consider two uniform, homogeneous and isotropic constituents, made of different materials. Even if one of the constituents were to be subjected to a constant temperature change before the assembly, the local symmetry groups would remain unchanged and the configuration would still remain homogenenous. On assembly and cooling, the composite remains uniform, homogeneous and isotropic. This is schematically illustrated in Figure \ref{fig:trivial}.
\begin{figure}[H]
\begin{center}
\begin{tikzpicture}[scale=0.5]
\draw[thick] (0,0)--(6,0)--(6,6)--(0,6)--cycle;
\foreach \x in {1,2,3,4,5}
\foreach \y in {1,2,3,4,5}
\draw (\x,\y) circle (0.4);
\begin{scope}[xshift=8cm]
\draw[thick] (0,0)--(6,0)--(6,6)--(0,6)--cycle;
\foreach \x in {1,2,3,4,5}
\foreach \y in {1,2,3,4,5}
\draw (\x,\y) circle (0.25);;
\end{scope}
\begin{scope}[xshift=16cm]

\draw[thick] (0,0)--(6,0)--(6,6)--(0,6)--cycle;
\foreach \x in {1,2,3,4,5}
\foreach \y in {1,2,3,4,5}
\draw (\x,\y) circle (0.32);
\end{scope}
\node at (7,3) {$+$};
\node at (15,3) {$=$};
\end{tikzpicture}
\end{center}
\caption{A locally trivial composite}
\label{fig:trivial}
\end{figure}
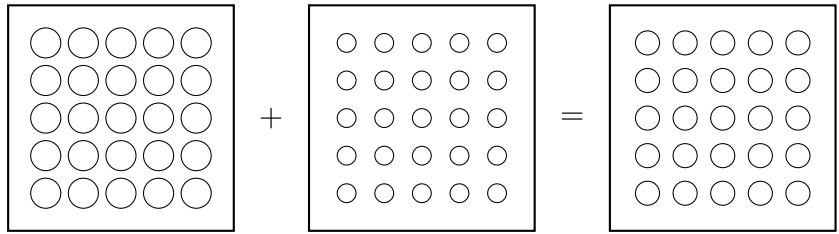
}
\end{example}

Finally, if the material double groupoid is only {\it horizontally locally trivial}, only the $\sqcup$ -shaped channels are guaranteed to have a closing side, which would imply that the vertex groups of the vertical side groupoid are subgroups of their horizontal counterparts. The binary composite will then mirror the uniformity of the vertical constituent. A similar reasoning can be applied to vertically locally trivial material double groupoids.

 If a binary composite turns out to be non-uniform, its material double groupoid may still have interesting features indicative of a physical property. A good example is that of a laminate, namely, a composite which is uniform along surfaces only. This feature would be reflected as properties of the transitivity components of the double groupoid, which can be defined in a manner analogous to the groupoid counterpart.

\begin{example} {\rm {\bf A laminated composite}:  An example of loss of uniformity with the emergence of a laminate is shown in Figure \ref{fig:laminate}.
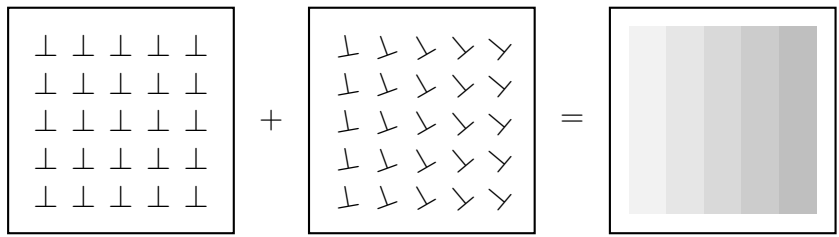
\begin{figure}[H]
\begin{center}
\begin{tikzpicture}[scale=0.5]
\draw[thick] (0,0)--(6,0)--(6,6)--(0,6)--cycle;
\foreach \x in {1,2,3,4,5}
\foreach \y in {1,2,3,4,5}
\node at (\x, \y) {$\perp$};
\begin{scope}[xshift=8cm]
\draw[thick] (0,0)--(6,0)--(6,6)--(0,6)--cycle;
\foreach \x in {10,20,30,40,50}
\foreach \y in {1,2,3,4,5}
\node at (0.1*\x,\y) {\rotatebox[origin=c]{\x}{$\perp$}};
\end{scope}
\begin{scope}[xshift=16cm]

\draw[thick] (0,0)--(6,0)--(6,6)--(0,6)--cycle;
\foreach \x in {10,20,30,40,50}
\path[fill=gray!\x] (0.1*\x-0.5,0.5)--(0.1*\x+0.5,0.5)--(0.1*\x+0.5,5.5)--(0.1*\x-0.5,5.5)--cycle;
\end{scope}
\node at (7,3) {$+$};
\node at (15,3) {$=$};
\end{tikzpicture}
\end{center}
\caption{Partial preservation of uniformity: a laminate}
\label{fig:laminate}
\end{figure}
}
\end{example}

\subsubsection{Isotropic constituents}

An isotropic elastic material point has, in any configuration, a symmetry group which is a conjugate of the special orthogonal group $SO(3)$.\footnote{Or of the full orthogonal group $O(3)$.} The material groupoid of a uniform body made of such a material is, accordingly, a transitive groupoid whose structure group can be chosen as any such group. In particular, it is possible to choose (as an {\it archetype}), the group $SO(3)$ itself as the structure group of the material groupoid.

Suppose that a binary composite is made of two constituents, each of which is a uniform isotropic body, and assume that the resulting composite turns out to be uniform. Since the structure group of each of the constituents can be chosen as $SO(3)$, it would appear that the structure group of the composite is also $SO(3)$. This wrong conclusion would stem from a careless reasoning based on the correct assertion that the commutator at each point of the composite must belong to the intersection of the local symmetry groups of the constituents.

Let the first constituent be a perfectly uniform and homogeneous body in a homogeneous configuration such that the local symmetry group $G_X$ at each point $X \in {\mathcal B}$ is, indeed, the special orthogonal group $SO(3)$. The second constituent is also uniform but not homogeneous. At each point $X \in {\mathcal B}$ the symmetry group ${\hat G}_X$ is a conjugate of $SO(3)$. In other words, there is a (smooth) field of non-singular matrices $H(X) \in GL(3;{\mathbb R})$ such that
\begin{equation} \label{dg13}
{\hat G}_X= H(X) G_X (H(X))^{-1}.
\end{equation}

The intersection $G_X \cap {\hat G}_X$ consists, therefore, of all orthogonal matrices in ${\hat G}_X$. Put differently, it consists of all matrices in ${\hat G}_X$ that are equal to their respective inverse transposes. From Equation (\ref{dg13}), it follows that this set consists precisely of all orthogonal matrices $Q$ satisfying the condition
\begin{equation} \label{dg14}
CQ=QC,
\end{equation}
where $C=H^TH$ is, at each point $X$, symmetric and positive definite. In the terminology of group theory, we can say that the intersection $G_X \cap {\hat G}_X$ is the {\it orthogonal normalizer} of $C$.\footnote{Since $C$ is a single element (rather than a general subset) of $ GL(3;{\mathbb R})$, the normalizer in this case coincides with the centralizer.}

Since, according to Equation (\ref{dg14}), the matrices $C$ and $Q$ commute, we know that the product $CQ$ preserves the eigenspaces of both factors. If the three eigenvalues of $G$ are distinct, the intersection consists of the unit matrix and the three rotations through the angle $\pi$ about each of the three eigendirections of $C$. The composite is {\it orthotropic}, corresponding to the {\it rhombic} system in the classification of crystals.

If two of the eigenvalues of $C$ coincide, the intersection  $G_X \cap {\hat G}_X$ is the group of rotations about the distinct eigendirection, and the material becomes {\it transversely isotropic}. Only when the three eigenvalues coincide, the composite becomes isotropic. In this case, each $H$ is a pure spherical dilatation, but, if the composite must be uniform, this dilatation must be a constant field, and we obtain a trivial case.

\begin{example} {\rm {\bf Isotropic constituents}: In the two-dimensional setting, the only two possibilities for a uniform binary composite generated by two uniform isotropic constituents are orthotropy and full isotropy. Let the first isotropic constituent be uniform and homogeneous, with a symmetry group everywhere equal to $SO(2)$. The second constituent is obtained via a conjugation of the form $H(X)=AQ(X)$, where $Q(X)$ is an orthogonal field, and $A$ is a fixed matrix. For arbitrary $Q(X)$, the resulting binary composite is uniform with a discrete symmetry group, as suggested by Figure \ref{fig:isotropic}.
\begin{figure}[H]
\begin{center}
\begin{tikzpicture}[scale=0.5]
\draw[thick] (0,0)--(6,0)--(6,6)--(0,6)--cycle;
\foreach \x in {1,2,3,4,5}
\foreach \y in {1,2,3,4,5}
\draw (\x,\y) circle (0.3);
\begin{scope}[xshift=8cm]
\draw[thick] (0,0)--(6,0)--(6,6)--(0,6)--cycle;
\foreach \x in {10,20,30,40,50}
\foreach \y in {1,2,3,4,5}
\draw[rotate around ={\x+7*\y+2*\y*\y:(0.1*\x,\y)}] (0.1*\x,\y) ellipse (0.2 and 0.4);
\end{scope}
\begin{scope}[xshift=16cm]

\draw[thick] (0,0)--(6,0)--(6,6)--(0,6)--cycle;
\foreach \x in {10,20,30,40,50}
\foreach \y in {1,2,3,4,5}
\draw[rotate around ={\x+7*\y+2*\y*\y:(0.1*\x,\y)}] (0.1*\x,\y-0.3)-- (0.1*\x,\y+0.3);
\end{scope}
\node at (7,3) {$+$};
\node at (15,3) {$=$};
\end{tikzpicture}
\end{center}
\caption{Two isotropic plates with loss of isotropy}
\label{fig:isotropic}
\end{figure}
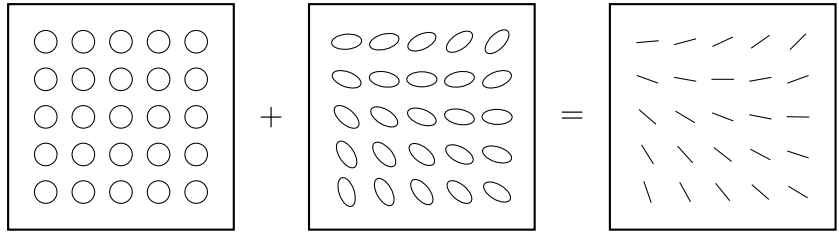
}
\end{example}
\section{Concluding remarks}

The material geometry of binary composites has been used to illustrate the potential of using the theory and formalism of double groupoids in continuum mechanics. It can be expected that this tool may open new possibilities for the description of objects of material geometry of a more complicated nature than that of binary composites. An apt way to conclude this article is perhaps to quote from the final remarks of Brown and Mackenzie \cite{brown1}:
\quote{In conclusion, it seems reasonable to expect that the classification of double groupoids will be more difficult, and exhibit a wider range of possibilities, than that for ordinary groupoids ... Perhaps we should not even expect there to be descriptions of {\it all} double groupoids in terms of other more familiar structures, but regard double groupoids themselves as basic objects in mathematics. On the other hand, where such descriptions are available, they can be of considerable use. }

\end{document}